\documentclass[a4paper,fleqn,usenatbib]{mnras}

\usepackage{newtxtext,newtxmath}

\usepackage[T1]{fontenc}
\usepackage{ae,aecompl}

\usepackage{natbib}
\usepackage{psfrag}
\usepackage{xspace}
\usepackage{color}
\usepackage{graphicx}   
\usepackage{amsmath}    
\usepackage{amssymb}    

\newcommand{\Teq}{$T_{\rm eq}$}
\newcommand{\Rpl}{$R_{\rm pl}$}
\newcommand{\Mpl}{$M_{\rm pl}$}
\newcommand{\Rer}{\ensuremath{R_{\oplus}}}
\newcommand{\Me}{\ensuremath{M_{\oplus}}}
\newcommand{\Msun}{M$_{\odot}$}

\newcommand{\fR}{$f_{\rm at} \times R_{\rm pl}$} 
\newcommand{\fat}{$f_{\rm at}$}
\newcommand{\fato}{$f_{{\rm at},0}$}

\title[Thermal evolution and the atmospheric escape]{Coupling thermal evolution of planets and hydrodynamic atmospheric escape in MESA}

\author[D. Kubyshkina et al.]{
Daria Kubyshkina,$^{1}$\thanks{E-mail: kubyshkd@tcd.ie} Aline
A.~Vidotto,$^{1}$ Luca Fossati$^{2}$ and Eoin Farrell$^{1}$
\\
$^{1}$School of Physics, Trinity College Dublin, the University of Dublin, College Green, Dublin-2, Ireland\\
$^{2}$Space Research Institute, Austrian Academy of Sciences, Schmiedlstrasse 6, A-8042 Graz, Austria\\
}

\date{Accepted XXX. Received YYY; in original form ZZZ}

\pubyear{2020}

\begin{document}
\label{firstpage}
\pagerange{\pageref{firstpage}--\pageref{lastpage}} \maketitle

\begin{abstract}
The long-term evolution of hydrogen-dominated atmospheres of
sub-Neptune-like planets is mostly controlled to by two factors: a
slow dissipation of the gravitational energy acquired at the
formation (known as thermal evolution) and atmospheric mass loss.
Here, we use MESA to self-consistently couple the thermal
evolution model of lower atmospheres with a realistic
hydrodynamical atmospheric evaporation prescription. To outline
the main features of such coupling, we simulate planets with a
range of core masses (5-20\Me ) and initial atmospheric mass
fractions (0.5-30\%), orbiting a solar-like star at 0.1~au. In
addition to our computed evolutionary tracks, we also study the
stability of planetary atmospheres, showing that the atmospheres
of light planets can be completely removed within 1~Gyr, and that
compact atmospheres have a better survival rate. From a detailed
comparison between our results and the output of the
previous-generation models, we show that coupling between thermal
evolution and atmospheric evaporation considerably affects the
thermal state of atmospheres for low-mass planets and,
consequently, changes the relationship between atmospheric mass
fraction and planetary parameters. We, therefore, conclude that
self-consistent consideration of the thermal evolution and
atmospheric evaporation is of crucial importance for evolutionary
modeling and a better characterization of planetary atmospheres.
From our simulations, we derive an analytical expression between
planetary radius and atmospheric mass fraction at different ages.
In particular, we find that, for a given observed planetary
radius, the predicted atmospheric mass fraction changes as
$age^{0.11}$.
\end{abstract}

\begin{keywords}
Hydrodynamics -- Planets and satellites: atmospheres -- Planets
and satellites: physical evolution
\end{keywords}



\section{Introduction}\label{sec::intro}

The evolution of planetary atmospheres is a complex topic, which
requires multiple considerations, including the formation of
planetary systems from the circumstellar disks
\citep[e.g.,][]{mizuno1980,morbidelli2009,morbidelli2016,morbidelli2020,mordasini2012,leconte2015},
the possible migration of planets
\citep[e.g.,][]{mordasini2015,morbidelli2016,morbidelli2020}, and
the atmospheric accretion
\citep[e.g.,][]{ikoma2012,steokl2015,steokl2016}. It then proceeds
with the thermal evolution of the atmosphere
\citep[e.g.,][]{rogers2010,nettelmann2011,miller2011,valencia2013,lopez2012,lopez2014},
the removal of  primordial atmospheres due to  atmospheric escape
processes
\citep[e.g.,][]{watson1981,lammer2003,lammer2016,Erkaev2007,erkaev2016,Lecavelier2004,salz2016,kubyshkina2018grid},
including the joint evolution of the host star
\citep[e.g.,][]{sanz2011,wright2011,Jackson2012,shkolnik2014,MCdonald2019}.
Accurately combining these ingredients is key for explaining the
existing population of exoplanets, as revealed by observations
\citep[e.g.][]{bonfils2013,mullally2015}, and unveiling the
history of planetary formation and their co-evolution with host
stars. As the evolution of planetary atmospheres is driven to a
large extent by stellar irradiation, the study of this evolution
can also provide the information needed to constrain the past
evolution of stars hosting planetary systems
\citep[][]{kubyshkina2019,kubyshkina2019b,owen2020}. In turn,
constraining the stellar evolution \citep[particularly at young
ages, when stars are known to follow various paths in terms of
their high energy luminosity, see e.g.][]{tu2015} is one of the
first and necessary steps in habitability studies
\citep[e.g.,][]{lammer2018,lammer2019}.

An accurate modelling of atmospheric evolution is also important
for characterisation of exoplanets and  interpretation of
observations. Recent missions, such as Transiting Exoplanet Survey
Satellite \citep[TESS;][]{ricker2015} and  CHaracterising
ExOPlanets Satellite (CHEOPS; \citealt{broeg2013}), are dedicated
to study  exoplanets and can provide an outstanding level of
measurement precision, which calls for finer analysis tools.

In this work, we concentrate on the thermal evolution of the
primordial atmospheres of sub-Neptune planets. For that, we use
MESA (Modules for Experiments in Stellar Astrophysics,
\citealt{paxton2018}), coupled with atmospheric mass loss. The
evolution of planetary atmospheres has been studied with MESA
before \citep[e.g.,][]{ikoma2012,lopez2012,lopez2013,steokl2016},
and the escaping atmosphere was modelled using an energy limited
approximation  \citep[e.g.,][]{watson1981,Erkaev2007}. This
approximation, however, has been shown to be inappropriate for the
study of sub-Neptune size planets
\citep[][]{steokl2016,lammer2016,owen2016,fossati2017,kubyshkina2018grid}.
{The novelty of our work here is that }we combine the study of the
thermal evolution with the more realistic prescription for
atmospheric mass loss from  hydrodynamical studies
\citep{kubyshkina2018grid,kubyshkina2018approx}. Our method allows
us to investigate how those processes interact with each other and
how much escape can affect planetary evolution.

An accurate prescription of atmospheric mass loss is of crucial
importance in the study of atmospheric evolution. A proper
coupling of the thermal evolution with atmospheric escape is
relevant for calculating the actual atmospheric mass fraction from
mass and radius measurements.

We consider sub-Neptune-like planets in the planetary mass range
of 5-20 Earth masses with initial atmospheric mass fractions
between 0.5 and 30\% orbiting a Solar-like star at the orbital
separation of 0.1~AU. These planets are highly affected by
atmospheric escape,  making them a good sample for investigation
of the possible effects of  coupling between atmospheric escape
and thermal evolution.

This paper is organised as follows. In Section \ref{sec::model} we
describe our modelling approach. In Section \ref{sec::comparison}
we present our results and compare how the evolution is affected
when assuming  different atmospheric mass loss prescriptions (the
simplified energy limited approach or the more robust hydrodynamic
prescription). Further, in Section~\ref{sec::fR}, we consider the
relation between atmospheric mass fraction and planetary
parameters throughout the evolution. We discuss our results and
give the conclusions in Section \ref{sec::discussion}.

\section{Modelling approach: coupling hydrodynamical escape simulations into MESA}\label{sec::model}

The `Modules for Experiments in Stellar Astrophysics' (MESA)
\citep{paxton2011,paxton2013,paxton2018} is a widely used tool in Stellar Physics, having a wide range of applications in the field of stellar evolution and internal structure {\citep[e.g.,][]{cantiello2014,choi2016,marchant2016,farrel2020a,farrel2020b}}. An interesting, but less common use of MESA, is that of modelling the thermal evolution of giant or sub-Neptune planets \citep[see e.g.][]{batygin2013,storch_lai2014,Nayakshin2015,jackson2016,Chen_rog2016,dederick2017,berardo2017,Chatterjee2018,malsky_rog2020}. Here we describe how we use  MESA to simulate the evolution of sub-Neptune planets. To describe such low-mass objects, 
MESA employs the equation of state (EOS) from \citet{saumon1995}
(SCVH EOS in MESA documentation) and use opacity tables from
\citet{freedman2008} \citep[further details on the use of MESA for
planetary modelling can be found in][]{paxton2013}.

\subsection{Initial conditions}\label{ssec::model-init}

In this work, we employ the MESA version 12115 \citep{paxton2018}
to model the thermal evolution of sub-Neptune planets in the mass
range between 5 and 20 Earth masses (\Me). To set up the planets,
which we further evolve, we follow the basic algorithm suggested
by \citet{Chen_rog2016}, which we describe here briefly.

\begin{enumerate}
    \item At the first step, we create a coreless planet of 0.1 Jupiter masses using \texttt{create\_initial\_model} (alternatively, one can use the initial model from the test suite available in MESA directory (\texttt{test\_suite/irradiated\_planet/0.001Msun.mod}) and reduce its mass using \texttt{relax\_initial\_mass}).
    \item Next, we insert an inert core\footnote{In the present paper, `core' is the solid part of the planet composed of silicates and metals, and we do not take into account its internal structure.} of the specific mass (5, 7.5, 10, 12.5, 15, and 20 \Me) using the \texttt{relax\_core} option, {hence modify the internal boundary condition.} Here, we adopt the bulk density of the core given by \citet{rogers2011core} for the rocky composition.
    \item We reduce the atmosphere to the desired volume using \texttt{relax\_initial\_mass}.  For each of the cores from the previous step we consider atmospheric mass fractions of 0.5, 1, 2, 3, 5, 10, 20, and 30\%.
    \item Further, we standardize the initial energy budget of the planet. We set an artificial luminosity with \texttt{relax\_initial\_L\_center} option as described in \citet{Chen_rog2016} to inflate the atmosphere, and then evolve the planet without atmospheric escape for 5~Myr, which is an average lifetime of protoplanetary discs \citep{mamajek2009}. 
    \item At the final step, we remove the artificial luminosity and set up an initial temperature at the upper boundary of 772 K using \texttt{relax\_initial\_irradiation} option. This temperature corresponds to the equilibrium temperature assuming zero albedo of a planet orbiting at 0.1 AU around a solar mass star at the age of 5 Myr \citep[resolved from the stellar parameters given by MESA Isochrones and Stellar Tracks; MIST][]{choi2016}.
\end{enumerate}

After these steps, the newborn planet is ready to evolve.%

\subsection{Thermal evolution and  atmospheric escape}\label{ssec:model-evolution}

To further account for the atmospheric heating by the host star
irradiation, we create a user routine using the
\texttt{other\_energy} module of MESA, which tracks the
equilibrium temperature of the atmosphere during the evolution as
resolved using MIST. This is different to the approach of
\citet{Chen_rog2016}, who assumed a constant temperature
(compatible with the present Sun) throughout the evolution of the
planet. We will discuss the implications from this update in the
next section. We also include in the heating function the
time-dependent luminosity of the core, accounting for its thermal
inertia and radiogenic heating \citep{Chen_rog2016}.

{The most important novelty of our work relates to the treatment
of atmospheric escape. For that, we use the results of
hydrodynamical (HD) simulations of atmospheric escape presented by
\citet{kubyshkina2018grid}. These models consider}  a pure
hydrogen atmosphere and describe atmospheric heating by absorption
of the stellar XUV (x-ray+extreme ultraviolet) flux; they account
for hydrogen dissociation, recombination, and ionization, and
$Ly\alpha-$\  and $H_3^+-$cooling. The grid of models in
\citet{kubyshkina2018grid} consists of about 7000 models and
covers planetary mass (\Mpl) in the range of $1-40$ \Me, planetary
radius (\Rpl) in the range of 1-10 \Rer, orbital separations
corresponding to equilibrium temperatures (\Teq) between 300 and
2000 $K$, stellar masses between 0.4 and 1.3 \Msun, and stellar
irradiation levels from the present Sun (scaled to the specific
stellar masses) to about $10^4$ times solar {(appropriate for
younger stars)}.

{Given that escape depends on the properties of the planet (mass,
radius, orbital separation) and of the star (age, XUV radiation),
it would have been computationally intensive to run one model for
each timestep of the MESA simulation. To overcome this,} we use
the semi-analytic approach described in
\citet{kubyshkina2018approx}. This analytical approximation is
based on the large grid of hydrodynamical 1D models of planetary
upper atmospheres {(described above)} and is very convenient for
incorporating in planetary evolution studies. The analytical
description has as arguments the planetary radius, mass and
temperature, orbital separation and stellar flux and is given by
\begin{eqnarray}\label{eqn::HBA}
\dot{M}_{\rm HBA} &=& e^{\beta}\,\,(F_{\rm XUV})^{\alpha_1}\,\,\left(\frac{d_0}{\rm AU}\right)^{\alpha_2}\,\,\left(\frac{R_{\rm pl}}{R_{\oplus}}\right)^{\alpha_3}\,\,\Lambda^{\kappa}\,, \\
\kappa &=& \zeta + \theta\,\,\ln\left(\frac{d_0}{\rm AU}\right)\,,
\end{eqnarray}

\noindent {where coefficients
$\alpha_1,\,\alpha_2,\,\alpha_3,\,\beta,\,\zeta,$ and $\theta$ are
given in Table~1 of \citet{kubyshkina2018approx}.} Hereafter, we
refer to this analytical approximation as the hydro-based
approximation (HBA).

{To compare the results of our novel model with previous work, we
also consider the energy limited approximation, in which the
escape rate is given as}
\begin{equation}\label{eq::energyLimited}
\dot{M}_{\rm EL} = \frac{\pi\eta R_{\rm pl}R_{\rm XUV}^2F_{\rm
XUV}}{GM_{\rm pl}K}\,,
\end{equation}
\noindent where the factor $K$ accounts for Roche-lobe effects
\citep{Erkaev2007} {and $\eta$ is the fraction of the incoming
energy that is used in the heating process}. To be consistent with
assumptions of the hydrodynamic models, we adopt here $\eta=15\%$.
Following \citet{Chen_rog2016}, we assume the absorption radius of
the stellar XUV radiation in Equation~\ref{eq::energyLimited} to
be
\begin{equation}\label{eq::Reuv}
    R_{\rm XUV} = R_{\rm pl} + H \ln\left(\frac{P_{\rm photo}}{P_{\rm
    XUV}}\right),
\end{equation}
where \Rpl\ is the photospheric radius of the planet, $H$ and
$P_{\rm photo}$ are the atmospheric scale height and the pressure
at the photosphere, respectively. The pressure at the XUV
absorption level is defined as $P_{\rm XUV} \approx (m_{\rm H}G
M_{\rm pl})/(\sigma_{\nu_0}R_{\rm pl}^2)$, where \Mpl\ is the mass
of the planet, G and $m_{\rm H}$ are the gravitational constant
and the mass of hydrogen, respectively, and $\sigma_{\nu_0} =
1.89\cdot 10^{-18} {\rm cm^{2}}$ is the absorption cross-section
of hydrogen defined for the typical energy of the XUV radiation of
20 eV. Despite Equation\,\ref{eq::Reuv} being a simplification,
the effective radii of XUV absorption defined this way are similar
to those obtained through HD simulations (see
Figure~\ref{fig::Reuv}).

To describe the time-dependent stellar flux  (erg s$^{-1}$
cm$^{-2}$) at the planetary orbit we use the power law
\begin{equation}\label{eq::Feuv}
    F_{\rm XUV} =
    \begin{cases}
    29.7 \left(\frac{\rm age}{\rm 1 Gyr}\right)^{-1.23}\left(\frac{d_0}{\rm AU}\right)^{-2}, \text{\,if\,} {\rm age} > t_{\rm sat}\\
    886\left(\frac{d_0}{\rm AU}\right)^{-2}\,, \text{\,if\,} {\rm age} \leqslant t_{\rm sat}
    \end{cases}
\end{equation}
where $d_0$ is the orbital separation of the planet and $t_{\rm
sat}$ is the time until which the star remains in the saturated
regime \citep[e.g.,][]{wright2011,tu2015,MCdonald2019}. We
introduced the latter, as we start our simulation at a very young
age, when the power law in the first line of
Equation~\ref{eq::Feuv} \citep{ribas2005} gives the
unrealistically high XUV levels. To avoid it, we set the
saturation level for $L_X/L_{\rm bol}$ at the typical for
near-solar mass stars value of $2\times10^{-4}$
\citep{MCdonald2019,Jackson2012}. Taking the average $L_{\rm bol}$
at first 100 Myr of evolution of $2.9\times10^{33}$ \citep[about
0.74 of the present Sun level; based on MIST][]{choi2016}, we
obtain $L_{\rm X}$ at the saturation stage of $5.8\times10^{29}$
erg/s, which we further convert into $F_{\rm XUV} =
8.86\times10^4$\,${\rm erg/(s\cdot cm^2)}$ at the orbital
separation of 0.1 AU \citep[see ][for the conversion
relation]{sanz2011}. The saturation time for the considered case
is therefore slightly less than 100~Myr.

The MESA inlists and supplement functions used in the present work
are publicly available in Zenodo Repository
(https://doi.org/10.5281/zenodo.4022393) \footnote{Using the
inlists available at the MESA market one may need to apply a
handful of minor corrections to comply with the most recent
releases of MESA. The present work is based on version 12115.}.

\section{Comparison between radiation hydrodynamics model and energy-limited approximation for atmospheric loss}\label{sec::comparison}

\subsection{The case of the 12.5~\Me\ core model}
\begin{figure}
  \includegraphics[width=\hsize]{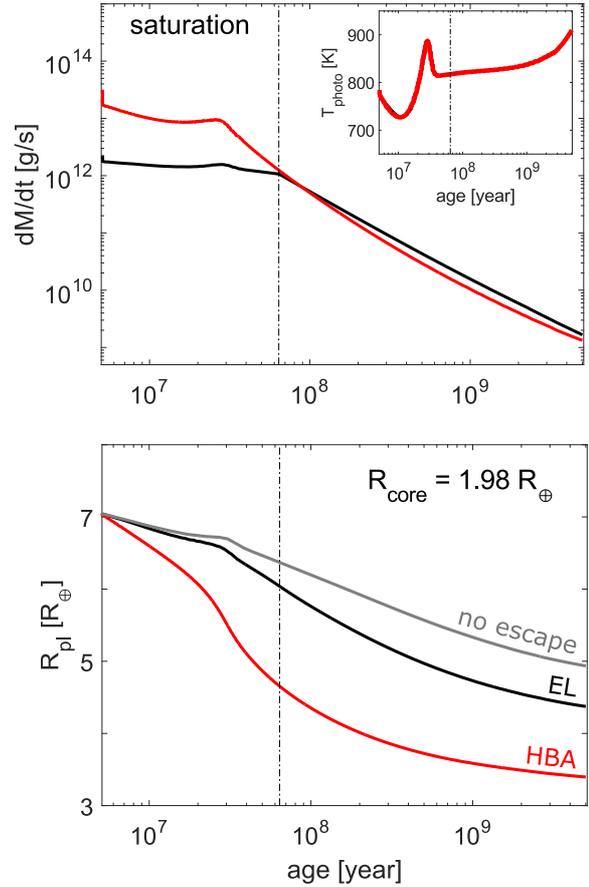}\\
  \caption{Evolutionary tracks of the atmospheric mass loss rate (top panel) and radius of the planet (bottom panel) considering different mass loss prescriptions: hydrodynamic models (HBA, red lines) and energy limited approximation (black lines).
  The gray line in the bottom panel shows how the radius of the planet changes in the absence of atmospheric escape, i.e., solely due to  thermal evolution. The vertical dashed-dotted lines denote the end of the period when the star remains in the saturation regime.
  With the additional sub-plot in the top panel we show how the photosphere {(equilibrium)} temperature changes with time to illustrate the adopted model of stellar heating.}\label{fig::evol}
\end{figure}

Before describing the results for the whole set of planets
considered in this work, we discuss a typical example of the
evolutionary tracks for the planet with a core mass of 12.5\,\Me\
and an initial atmospheric mass fraction of 20\% {(the total mass
of the planet at the beginning of the simulation is therefore
$M_{\rm pl,0} = M_{\rm core}/(1-\frac{f_{\rm at,0}}{100\%}) =
15.63$\,\Me)}, to introduce some of the main features of our
results. In Figure~\ref{fig::evol} (top panel), we present the
atmospheric escape rates ($\dot{M}$) throughout the evolution for
HBA (red line) and energy-limited (black) prescriptions, denoting
the duration of the saturation period of the host star {by the
vertical dashed-dotted line}. One can see, that two mass loss
mechanisms give significantly different outputs at the beginning
of the {evolution}, when the atmosphere of the planet is hot and
inflated. This period is relatively short: {about 60 Myr for the
considered case}. Afterwards, the two mass loss prescriptions
predict similar escape rates. This means, that most of the effect
to the final mass-radius distribution caused by the choice of the
mass loss prescription is decided during this short period after
the protoplanetary disk dispersal. We note that the duration of
this period does not always coincide with the saturation time as
it happens for the planet in Figure~\ref{fig::evol}; {this period}
is shorter for the lighter planets, as their atmospheres are
experiencing stronger escape, and their radii are therefore
decreasing fast, thus increasing the planetary gravity. This
higher gravity reduces the mass loss rate, thus extending the
timescale for the survival of the atmosphere. The differences in
mass loss rates are in general higher for lighter planets,
including for the later stages in evolution.

In the bottom panel of Figure~\ref{fig::evol} we present the
evolution of the planetary radius, including the case of a purely
thermal evolution in the absence of atmospheric escape (gray
line). The shape of the evolutionary track in the energy limited
case (black line) is mainly governed by the thermal evolution of
the planet, with atmospheric escape gradually decreasing the
radius of the planet. The bump at $\sim$30~Myr pairs with the
heating of the atmosphere shown in the equilibrium temperature
profile in the inset plot in the top panel. {As we can see in
Figure~\ref{fig::evol}, the final radii of the planet at 5Gyr is
very different depending on the assumption made for the
atmospheric escape. While the energy limit is an improvement
related to the case where no escape is considered, the models
using the energy-limited approximation still over-predict the size
of the planet, compared to the more realistic escape model using
radiation hydrodynamics.}

{As we noted in the introduction, one additional difference
between our models and those by \citet{Chen_rog2016} lies on the
treatment of the temperature of the planet. While in their models
they assume the planet remains at the same equilibrium temperature
throughout the evolution, in our models, we use a stellar
evolutionary track to derive the stellar bolometric luminosity and
thus the equilibrium temperature of the planet (inset in the top
panel of Figure~\ref{fig::evol}). We find that} the evolutionary
tracks in case of the energy limited atmospheric escape are nearly
unaffected by considering a time-dependent equilibrium
temperature. However,  for the HBA model, the difference can be
significant -- up to $\sim$20\% in terms of radii for the
considered set of planets. This can be explained by the fact that
the equilibrium (photospheric) temperature is not part of
Equation~\ref{eq::energyLimited} and affects the $\dot{M}_{\rm
EL}$ indirectly only through the variation of the planetary
radius. In the case of the HBA models (Equation~\ref{eqn::HBA}),
this temperature is accounted for through the orbital separation
\citep[for details
see][]{kubyshkina2018grid,kubyshkina2018approx}, and the
gravitational parameter of the planet $\Lambda$.

\subsection{Evolution for different core masses {and initial atmosphere sizes}}
\begin{figure}
  \includegraphics[width=0.9\hsize]{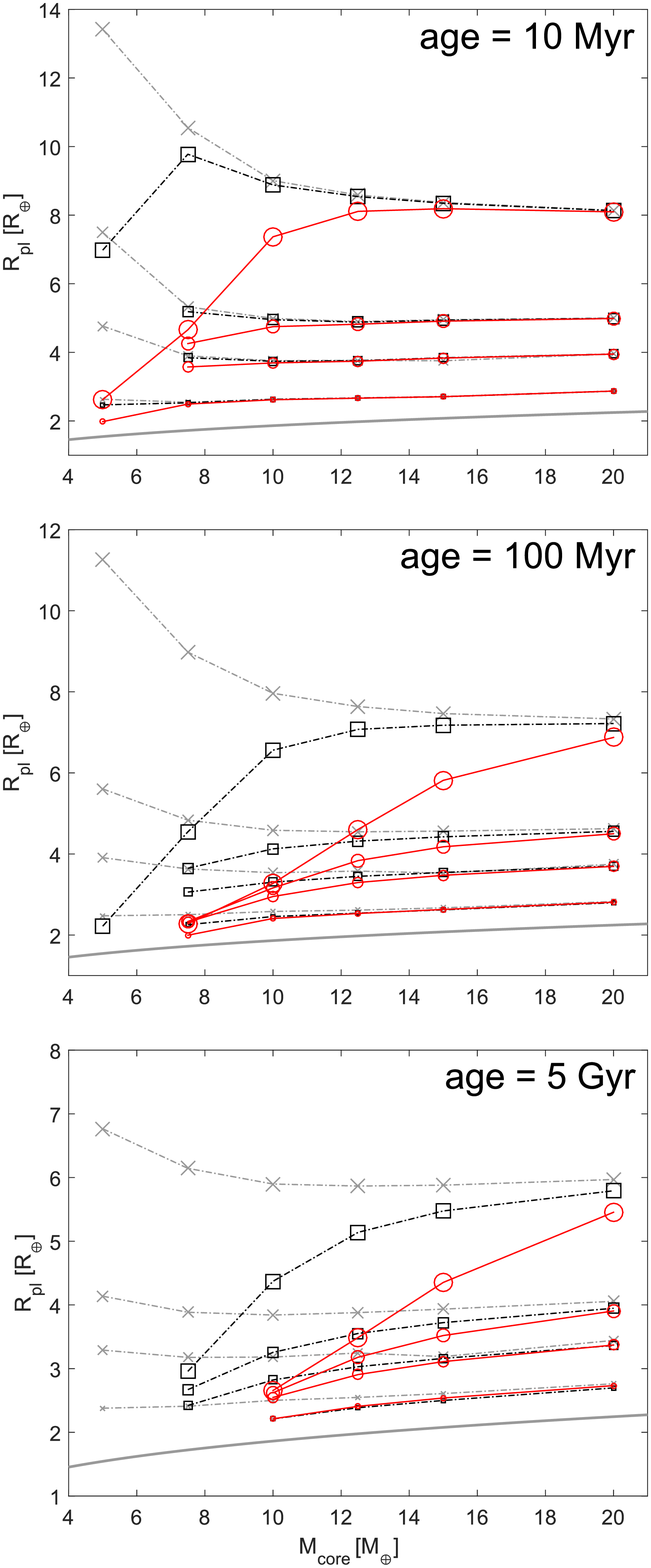} \\
  \caption{Mass-radius diagrams throughout the evolution at ages 10~Myr, 100~Myr, and 5~Gyr (from top to bottom panels). Black lines with square symbols correspond to the energy-limited atmospheric mass loss, red lines with circles to the hydrodynamic prescription (HBA), and the light gray lines with crosses correspond to the thermal evolution of the planet without mass loss. The size of symbols increases with initial atmospheric mass fractions, from  bottom to  top curves: 1, 5, 10, and 30\%. Note that some of the points are omitted at the low mass range at 100~Myr and 5~Gyr due to the complete loss of the atmosphere. {The gray solid lines represent the core radii accepted for the different planetary masses in this study \citep[][]{rogers2011core}}}\label{fig::panels}
\end{figure}

As a result of what shown above, for different atmospheric mass
loss prescriptions, the distributions of atmospheric mass
fractions and planetary radii against planetary core masses look
different throughout the evolution. In Figure~\ref{fig::panels},
we show the mass-radius diagram of the planet at the ages of
10~Myr, 100~Myr, and 5~Gyr. Here, the red solid lines correspond
to the HBA mass loss prescription, and the black dashed lines
correspond to the energy-limited prescription; lines with
different symbol sizes correspond to the different initial
atmospheric mass fractions (from 1\% to 30\% for lines from top to
bottom within each panel). Similar to what was shown in the
example above, the mass loss rates for these two prescriptions
differ dramatically in the first tens of Myr for the lighter
planets in the sample to $\sim$100~Myr for the heavier ones; the
difference itself is particularly large for lower gravity and
highly inflated planets (i.e., those having a large initial
envelope, see the case of 30\% initial atmosphere). Despite this
period being relatively short, it basically defines the final
state of the planet, and therefore the difference in planetary
radii predicted by the two different mass loss prescriptions
remains nearly the same from 100~Myr to 5~Gyr. {One can notice
that some of the symbols present in the top panel (10~Myr) are
missing in the two bottom panels for the cases of escaping
atmospheres. This occurs due to the complete loss of the
atmosphere taking place before 100~Myr (see, e.g., the 5\,\Me\
planet for the HBA prescription in red) or before 5~Gyr (e.g., the
7.5\,\Me\ planet and HBA atmospheric mass loss).}

{The third set of curves we show in Figure ~\ref{fig::panels} is
given by the grey lines - these curves represent the cases where
no escape is considered throughout the evolution of the planet.
This means that the size of the planet evolves with time only due
to the thermal evolution of their atmospheres. Similarly to what
we exemplified in Figure~\ref{fig::evol},  not considering escape
over-predicts the final sizes of the planets. }

Considering the HBA mass loss prescription {(red curves in
Figure~\ref{fig::panels})}, one can see that {the planets with
core masses below 10\,\Me\ end up with the same radii}
independently of the initial size of their atmospheres {(except
for the most compact initial atmospheres, see the case of $f_{\rm
at,0} = 1\%$)}. Although these planets start their evolution with
different radii, depending on their initial \fat , after a few
tens of Myr, we see that their `final' radii converged to the same
value and {changed only slightly for the rest of their lifetime.}
This effect is caused by the following: when increasing the amount
of atmosphere for a low-mass planet {at the formation stage}, the
radius increases significantly (this is seen well at the radius
distribution in absence of mass loss, grey lines in
Figure~\ref{fig::panels}{, top panel}), while the total mass of
the planet changes only slightly, {as the atmosphere is the minor
part of the total mass of the planet}. The HBA atmospheric escape
rate depends on the planetary radius to more than a power of 3 in
the ``low gravity'' regime (which is usually the case for the
young planets). As a result, after the amount of atmosphere
reaches some critical value {(which is $\sim$5\% and $\sim$10\%
for the planets of 7.5 and 10\,\Me core masses considered here,
respectively)}, the further increase in radius leads to a
consequent increase of the atmospheric mass loss rate. This effect
was considered in details in \citet{kubyshkina2019}.

\subsection{Survival timescales of atmospheres}\label{ssec::tau}
\begin{figure}
  \includegraphics[width=\hsize]{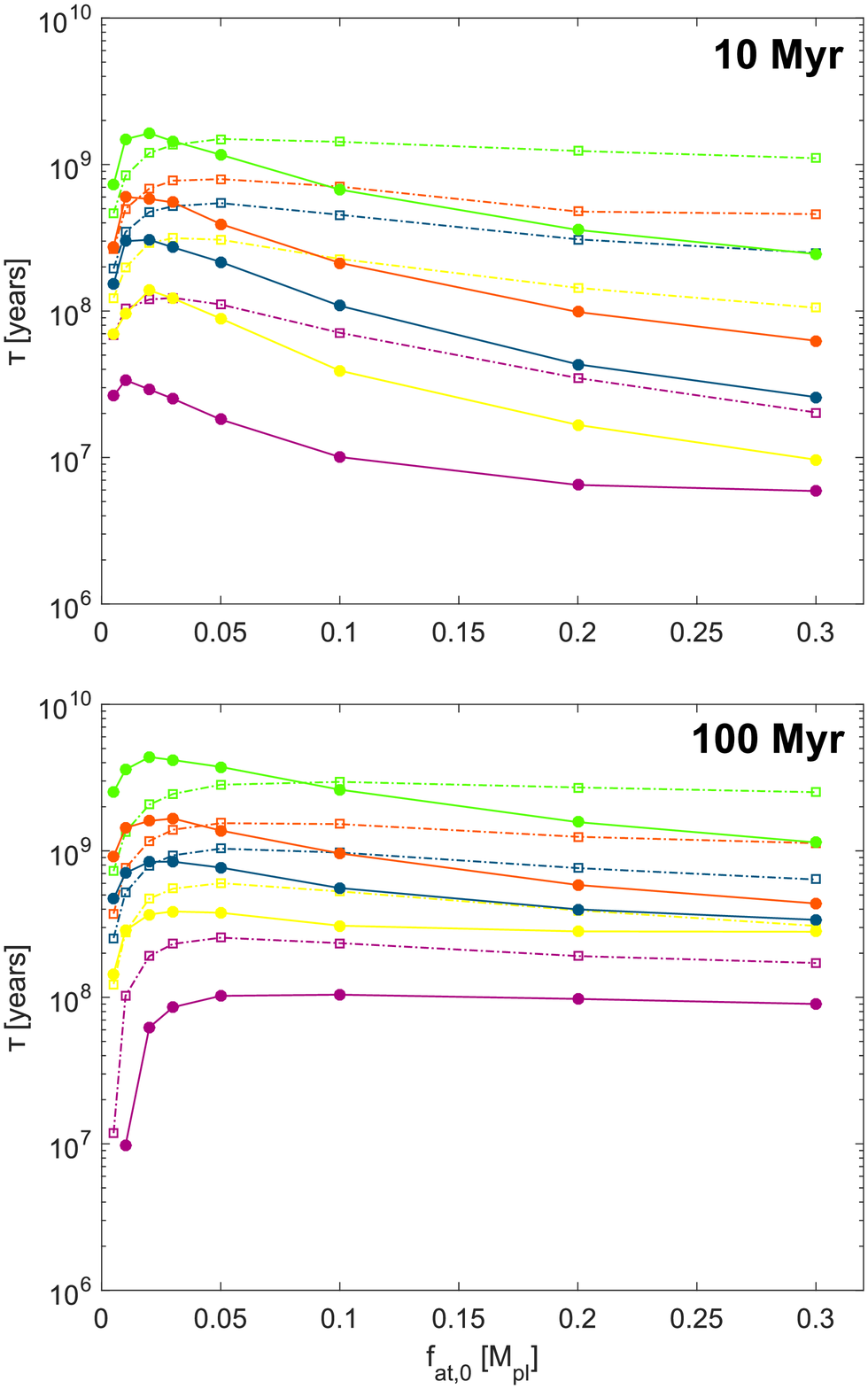} \\
  \caption{Instantaneous timescale for survival of atmospheres as a function of the initial atmospheric mass fraction at 10 and 100\,Myr. Solid lines correspond to the hydrodynamic mass loss, and dashed lines to the energy-limited mass loss. Different colors denote different planetary core masses -- from  bottom to  top, they are: 7.5, 10, 12.5, 15, 20\,\Me.}\label{fig::tau}
\end{figure}

In Figure~\ref{fig::tau}, we present the atmospheric escape
timescale $\tau$ at 10 and 100 Myr against the initial atmospheric
mass fraction \fat$_{,0}$. This parameter $$\tau = M_{\rm
atm}/\dot{M}$$ represents the instantaneous time that is needed to
fully remove an atmosphere of  mass $M_{\rm atm}$ if the mass loss
rate $\dot{M}$ remained constant. The larger the value of $\tau$,
the larger is the survival chance of the atmosphere. As most of
the atmospheric losses occur at the early stages of evolution, we
consider the parameter $\tau$ for 10 and 100~Myr only in
Figure~\ref{fig::tau}. The different line colors in
Figure~\ref{fig::tau} represent different masses of planetary
cores (increasing from the bottom to the top); solid lines
correspond to the HBA prescription, and dashed lines correspond to
the energy-limited prescription.

Both mass loss prescriptions predict similar dependencies: the
longest timescales peaks at relatively compact initial envelopes.
However,  the survival chance decreases steeply towards the
smallest atmospheres, and it also decreases towards the largest
one, because at some point the increase in escape rate overcomes
the increase in the mass of the atmosphere. Nevertheless, a closer
analysis reveals some differences between the two escape
approaches. The HBA {predicts larger survival timescales for} the
smaller  envelopes compared to the energy-limited approximation
($\tau$ maximizes at \fato $\sim 1-2$\% compared to $3-10${\% for
the energy limited approach}). Also, for HBA, the maximum is more
clearly pronounced and the shape of the curve remains similar for
all planetary core masses in the considered interval, while for
the energy-limited approach the maximum in $\tau$ is more
pronounced for the lighter planets. The only exception for the
shape of $\tau(f_{\rm at,0})$-dependence in case of HBA is the
flattening of the curves at low masses and large initial
envelopes, which is caused by the existence of the ``critical''
size of the atmosphere discussed above {(which is about 5\% for
the 7.5\,\Me\ planet, and about 10\% for the 10\,\Me\ planet
considered here)}.

\begin{figure}
  \includegraphics[width=\hsize]{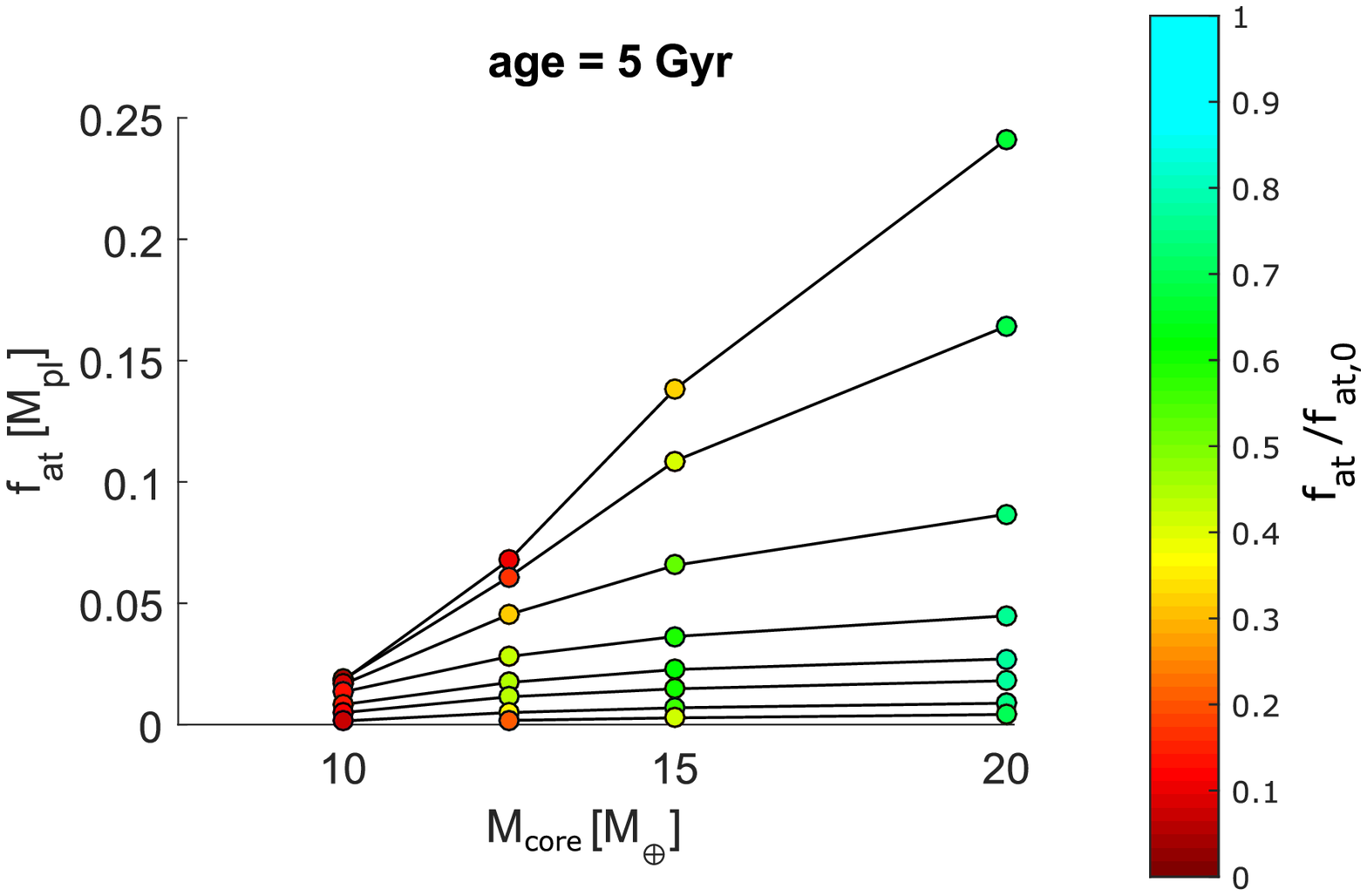}
  \caption{Atmospheric mass fractions against planetary (core) masses throughout the evolution assuming the HBA mass loss prescription. The different lines correspond to the different initial atmospheric mass fractions, from bottom to top: 0.5, 1, 2, 3, 5, 10, 20, 30\%. The color code shows the fraction of initial atmosphere preserved at 
  5~Gyr
  .}\label{fig::fat_eoin}
\end{figure}

{To further illustrate an increase in survival chances for the
mid-compact atmospheres, we show in Figure~\ref{fig::fat_eoin} the
distribution of the atmospheric mass fractions against the
planetary core masses of our simulated planets at the age of
5~Gyr. The different lines represent different initial atmospheric
mass fractions; the color code shows the relation of instant
atmospheric mass (at 5~Gyr) to the initial one ($\frac{f_{\rm
at}}{f_{\rm at,0}}$). For all considered core masses, the
atmospheres started as compact ones preserve the {largest} part of
initial masses compare to very thin or highly inflated ones.}

{For 10\,\Me\ core, however, not more than 10\% of initial
envelope is preserved in any of the cases. This makes a lucky
combination of the parameters important for the survival of the
atmosphere around such planet if it was exposed to the stronger
irradiation from the host star (e.g. if it would orbit closer
in).}

\section{Relation between planetary radii and atmospheric mass fraction}\label{sec::fR}

{Through observations, we can constrain the radii and masses of
exoplanets. Ideally, we would like to find a relationship between
\{\Mpl , \Rpl\} and atmospheric mass fraction from our models.
Therefore, once \{\Mpl , \Rpl\} are  observationally derived
(provided we know the system's age), we can infer some physical
properties for the planetary atmosphere. In this section, we
explore the possibility of finding a relation between planetary
radii and atmospheric mass fraction  (hereafter, \fR -relation). }

\subsection{Basic trends in thermal evolution}\label{sec::fR_trends}

\begin{figure*}
  \includegraphics[width=\hsize]{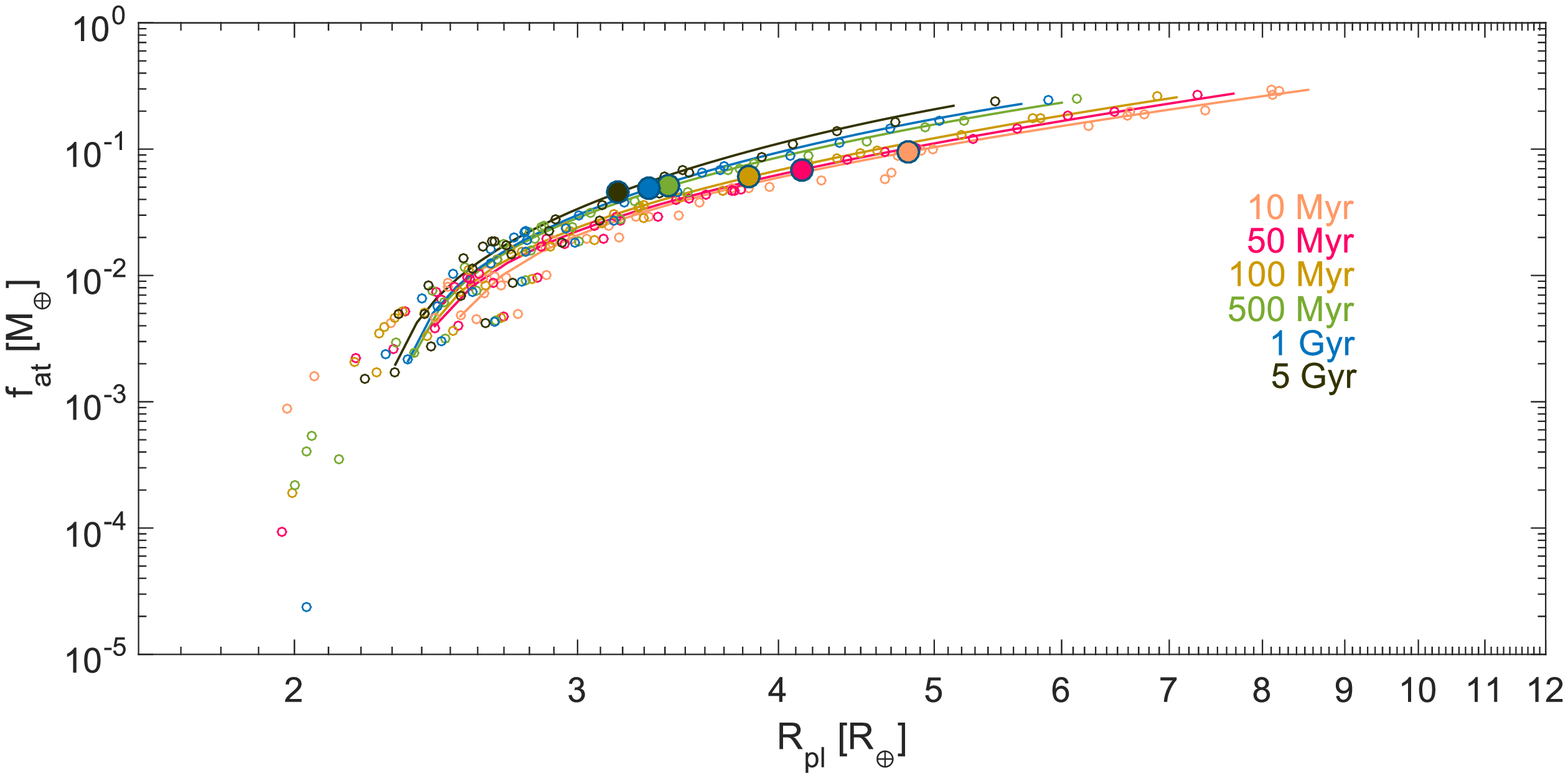}\\
  \caption{Evolution of the relation between \fat\ and \Rpl\ for all modeled planets. Small symbols represent the instantaneous radius and atmospheric mass fraction of each individual planet at a specific age, which are color coded as given in the legend.
  Big circles show the evolution of the \fR -relation with time for the planet with 12.5\Me\ core and \fato = 10\%. The {dashed} lines are the cubic polynomial fits of the \fR -relation for the planets with 12.5\Me\ core and initial atmospheric mass fractions between 0.5 and 30\% at different times.}\label{fig::rpl_fats}
\end{figure*}

{In this Section, we aim to study the behavior of the \fR
-relation depending on the system parameters (as age and planetary
mass) for the given set of simulated planets.}
Figure~\ref{fig::rpl_fats} shows with  small symbols all the
planets we modeled with ages between 10~Myr and 5~Gyr (color
coded). It is interesting to see that at each age, {and for each
considered mass} these points lie roughly along one line, which is
well fitted by a cubic polynomial (solid lines in
Figure~\ref{fig::rpl_fats}). {We give the coefficients of the
cubic fits in the Appendix~\ref{apx::cfu}}.

We note from Figure~\ref{fig::rpl_fats} that  the value of \fat\
for a specific radius is larger for older systems. We can also see
that the line curvature increases, {becoming more convex at older
ages}. From this, we infer that the shape of this line is
controlled by the thermal evolution of the planet (including the
input from stellar heating).
For each planetary mass (with various initial atmosphere masses) this line can be approximated by a cubic polynomial with a mean squared error of $10^{-6}-10^{-7}$. 
{This can be useful, for example, when one derives the radius of
the planet from observations and would like to estimate its
atmospheric mass fraction.}
Therefore, the evolution of the \fR -relation of an individual planet can be considered as the sum of two processes: change of the \fR\  curve with time due to thermal evolution, and the movement of the planet along this curve towards smaller radii due to  atmospheric escape, where the speed of this movement is defined by the mass loss prescription. We illustrate this process in Figure~\ref{fig::rpl_fats} for the 12.5\,\Me\ planet with \fato =  10\%: the position of the planet at each moment in time is shown by the large circles. 

\begin{figure}
  \includegraphics[width=\hsize]{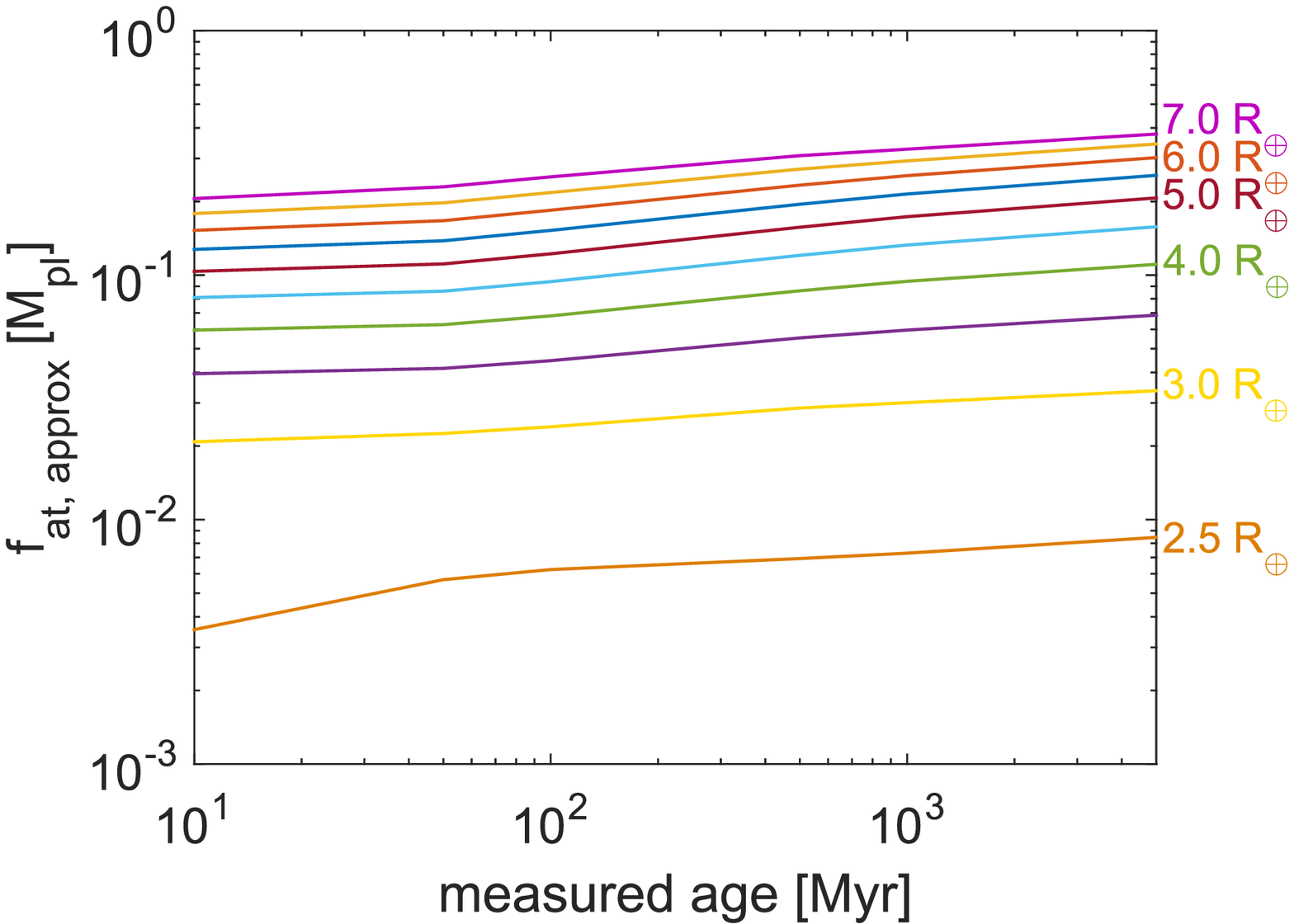}\\
  \includegraphics[width=\hsize]{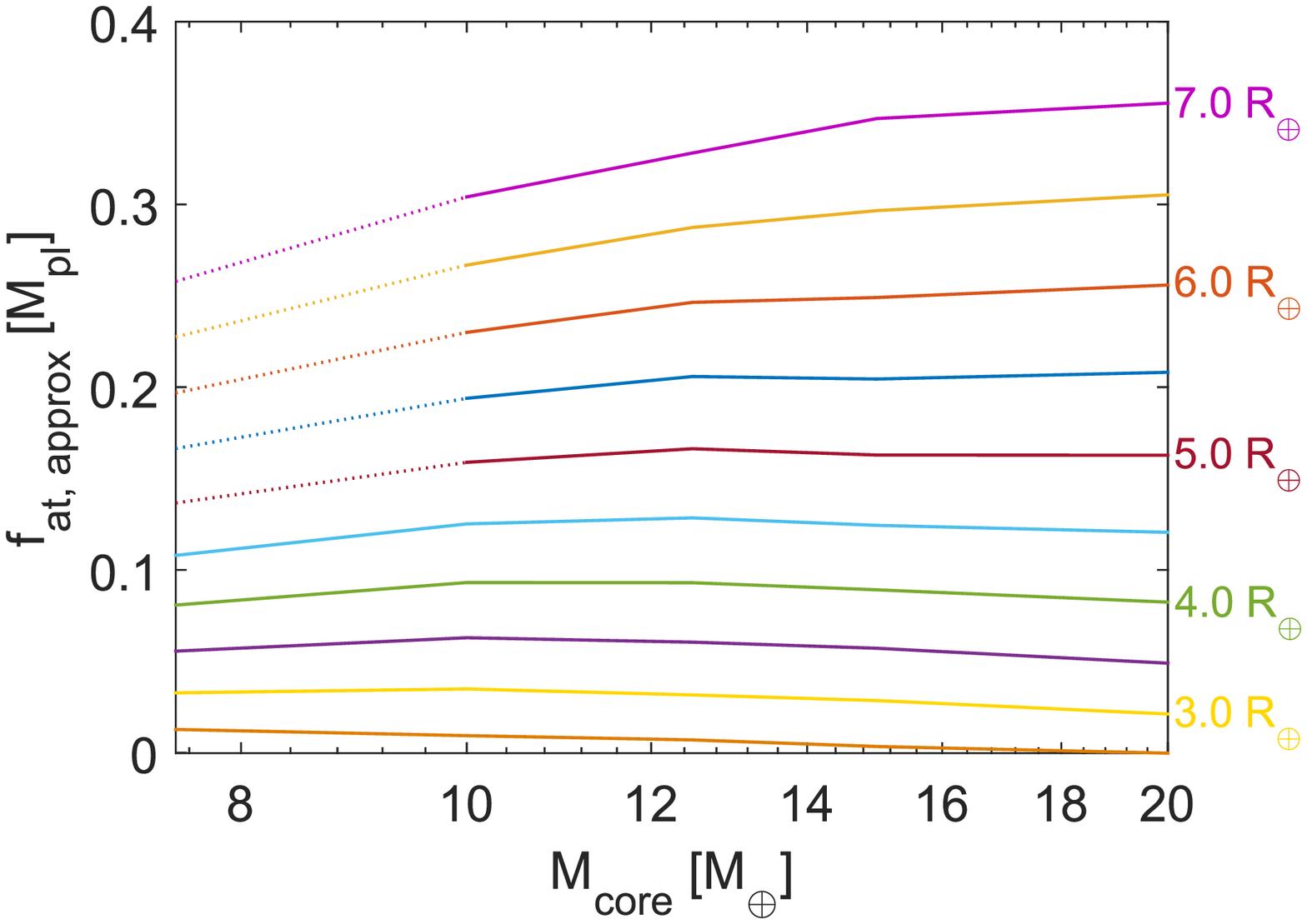}\\
  \caption{Atmospheric mass fraction as a function of age (of the system hosting a 12.5 \Me\ planet, top) and core mass (bottom) obtained employing the cubic polynomial approximation of the \fR -relation. Line colors correspond to planetary radii between 2.5 and 7\,\Rer with a step of 0.5\,\Rer (from bottom to top).
  In the bottom panel, the lines are dotted where the number of points used in the approximation is reduced (mass dependence is resolved only for ages < 500~Myr).}\label{fig::fat_time_12.5}
\end{figure}

In Figure~\ref{fig::fat_time_12.5} (top panel) we show how the
cubic polynomial approximation of \fR\ we obtained for the
12.5\,\Me\ planet changes with age for different radii between 2.5
and 7\,\Rer\ (from bottom to top). Each of the considered radii
corresponds to smaller \fat\ at earlier ages, due to higher
temperatures. One can see that $f_{\rm at, approx}$(age) is nearly
linear in the logarithm space; it means we can estimate the power
law that describes this relation. Averaging among the different
masses in our data set, we obtain
\begin{equation}\label{eq::fatage}
    f_{\rm at, approx}\propto \left(\frac{\rm age}{\rm 1 Gyr}\right)^{0.11},
\end{equation}
for each considered radii {and where the proportionality constant
is different for each \Rpl }. Equation~\ref{eq::fatage} remains
valid also in case of non-escaping atmospheres, which means that
this trend is general for the thermal evolution and is weakly
affected by differences in the thermal state of the planet caused
by the escape  {(although the proportionality constant changes for
different escape mechanisms)}.

The dependence of the \fat\ given by the cubic polynomial
approximation on the planetary mass (see
Figure~\ref{fig::fat_time_12.5}, bottom panel) is less simple. For
the lowest atmospheric mass fractions, the radius of the planet
(which can be considered as a sum of core radius and thickness of
the atmosphere) depends strongly on the mass-dependent core size
of the planet. This results in a negative correlation between
\fat\ and  planetary mass (\fat\ $ \propto M_{\rm pl}^{-0.5}$) for
\fat\ $\lesssim$10\%. For  \fat\ $\sim15\%$,  \fR\ is nearly
constant with mass, while for larger \fat\ $\gtrsim15\%$, we see a
positive trend with mass, changing from $\propto M_{\rm pl}^{0.1}$
to $\propto M_{\rm pl}^{0.25}$, for \fat\ from 15 to 30\%,
respectively. This difference in $f_{\rm at, approx}(M_{\rm pl})$
dependence reflects the change in the thermal evolution of the
planet due to  atmospheric escape: since the thermal state of low
mass planets is the most affected  (while the heaviest planet in
the simulation set is nearly not affected), the contrast in the
\fR -relation between low and high planetary masses increases with
inclusion of atmospheric mass loss. This confirms our inference
that the self-consistent inclusion of the thermal evolution and
realistic atmospheric escape is relevant for the evolution of
low-mass planets.

\subsection{Comparison of the \fR -relation with models that do not self-consistently account for escape and thermal evolution}\label{sec::fR_comparison}

\citet{lopez2014}  showed that, due to the thermal evolution of
the atmosphere, the relation between  atmospheric mass fraction
and  planetary radius can not be uniquely defined from the
measurable parameters of the planet, such as planetary mass and
equilibrium temperature. In fact, the \fR -relation also depends
on the age of the planet (i.e., on the current state of its
thermal evolution), and the dependence on age is stronger than the
dependence on the incident {XUV} flux from the star. The inclusion
of atmospheric escape can complicate the situation even more, as
escape is coupled with thermal evolution: loss of the atmosphere
speeds up the cooling of the planet, and, in turn, the cooling of
the planet and consequent shrinking the radius reduce mass loss.
From our simulations, the difference in temperature at the bottom
of the atmosphere after 5~Gyr reaches up to 3000~K, when comparing
cases with HBA escape and when escape is neglected. The largest
temperature differences are for the smallest planetary masses and
largest atmospheric mass fractions, while for heavy planets with
compact atmospheres they can be insignificant (e.g., for the
20\,\Me\ planet and \fato $\lesssim 10\%$ the difference in
temperature at the bottom of the atmosphere does not exceed
200~K).

\begin{figure*}
  \includegraphics[width=\hsize]{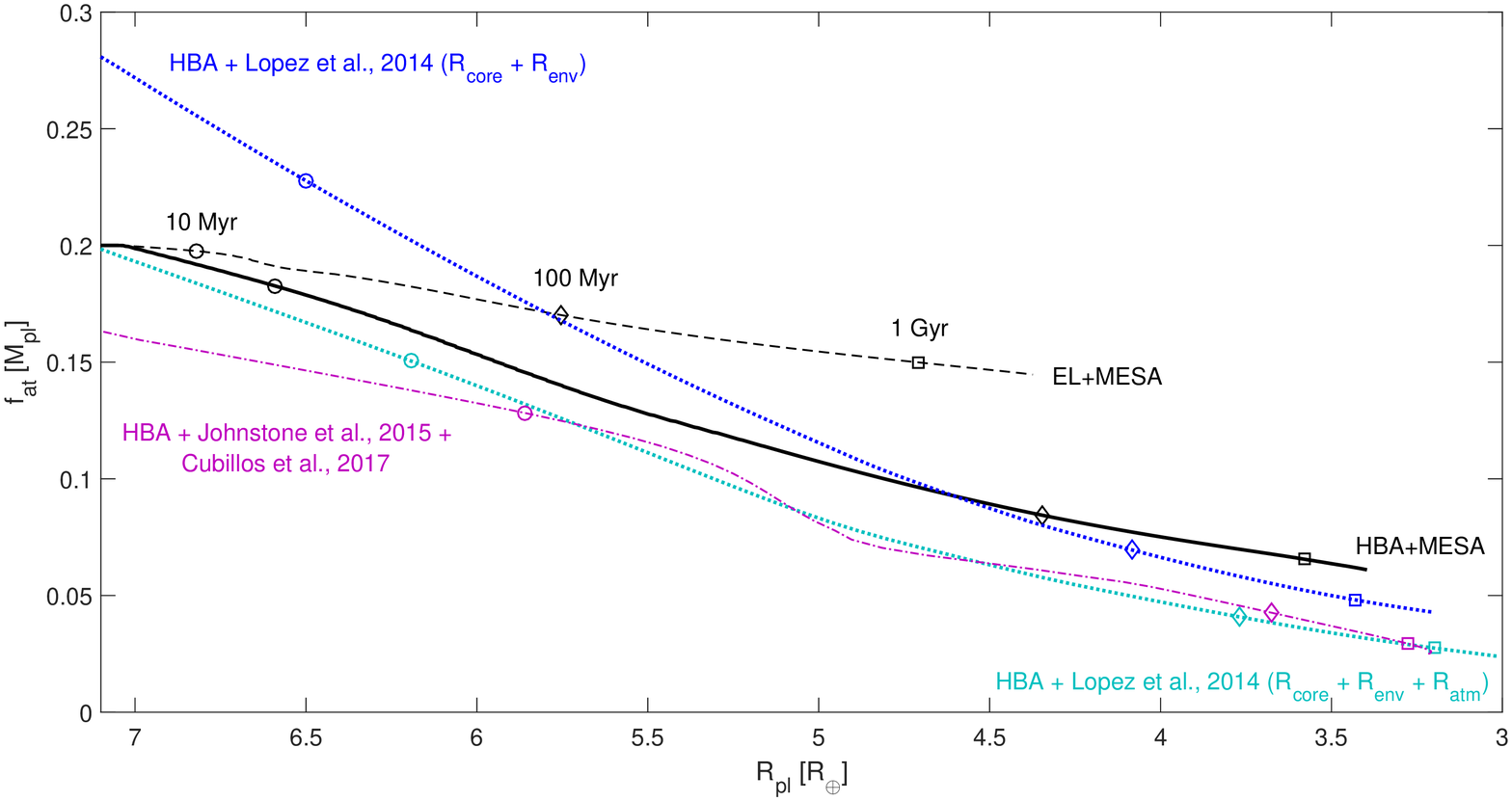}\\
  \caption{Relation between  atmospheric mass fraction and  planetary radius throughout the evolution for a planet with a core mass of 12.5\,\Me\  and 20\% of initial atmospheric mass fraction. Black lines represent the result given by the self-consistent modelling of thermal evolution of the planet (MESA) and atmospheric mass loss; solid line corresponds to the hydro-based approximation (HBA) mass loss prescription and dashed line corresponds to energy-limited (EL) prescription.
  {The colored lines represent the evolutionary tracks that do not take into account the self-consistent modelling of atmospheric escape and thermal evolution. These models employ semi-analytical approximations {to define planetary radius
  }. The light blue dotted line considers  the analytical approximation of \citet{lopez2014}, the dark blue dotted line uses the same approximation, but  excludes the upper radiative part of the atmosphere, and the violet dashed-dotted line shows the approximation based on \citet{johnstone2015}. For these three lines, we compute at each instant the escape of the planetary atmosphere using the HBA prescription -- {the radius is then recalculated for the current \fat\ }using the semi-analytical approximations. Because this is not self-consistent, the structure of the planet does not `respond' to mass loss in the same way as in our MESA models.}
}\label{fig::rpl_fats_intro}
\end{figure*}

In Figure~\ref{fig::rpl_fats_intro}, we consider how the \fR
-relation  changes due to mass loss for the case of the planet
with the core mass of 12.5~\Me, and with the initial atmospheric
mass fraction of \fato = 20\%. Here, the black solid line
represents the relation obtained in the present framework, i.e.,
MESA combined with HBA prescription for the mass loss. The black
dashed line shows the evolution using MESA, but now assuming the
energy limited mass loss mechanisms. One can see, that despite the
same starting point, the two tracks look different, and therefore
if, for the  energy-limited approach an atmosphere of \fat
$\sim$15\% is reached at an age of 5~Gyr when the radius is
$\sim$4.4\Rer, for HBA the same atmospheric mass fraction is
reached at an age below 100~Myr, when the radius of the planet is
$\sim$6\Rer .

Therefore, similarly to what was concluded by \citet{lopez2014},
{these different radii} for a specific amount of atmosphere are
achieved at different moments in the evolution and, thus, at
different thermal states of the atmosphere. From the observer's
point of view, this means that by measuring the radius of the
planet, without knowledge of its age, one can not derive the
amount of atmosphere the planet has.

\subsubsection{Comparison using the 12.5\Me -core planet}

{In addition to the faster/slower transition between thermal
states due to the specific atmospheric escape mechanism, the
thermal state at each given time can be altered due to the
coupling of thermal evolution with atmospheric mass loss.} In
order to separate and quantify the effects of these two processes
individually, we consider three additional \fR\ tracks. For these
tracks, we couple the HBA prescription of atmospheric mass loss to
three semi-analytical prescriptions between atmospheric mass
fraction and planetary radius that have been adopted in the
literature. We assume starting parameters of the planet equivalent
to those with which we start the MESA run. Next, we calculate the
atmospheric mass loss rate with HBA, define the appropriate time
step (requiring not more than 0.5\% of the whole atmosphere to be
lost within one step), and infer the atmospheric mass accordingly.
Then we estimate the new radius of the planet from the new
atmospheric mass fraction and adjusted parameters of the system.
We repeat this process until we reach the final time of 5~Gyr or
the atmosphere is totally gone. We vary the incident flux from the
star and the temperature at the photosphere throughout the
evolution in the same way as for MESA runs. The detailed
description of this approach can be found in
\citet{kubyshkina2018grid,kubyshkina2019}. {Note that, contrary to
our study performed with MESA in which escape is self-consistently
coupled to the thermal evolution of planets, the use of
semi-analytical \fR -relations plus the prescription of escape is
not self-consistent.}

As a first \fR\ prescription, we employ the analytical
approximation given in \citet{lopez2014}. Their model considers
the thermal evolution of the planetary atmosphere in absence of
mass loss and employs a similar approach as MESA, including the
same EOS and opacity tables; the time-dependent luminosity of the
core in this model follows the approach of \citet{lopez2012} and
is the same as in the present work. Most of the difference with
respect to the approach in this paper comes therefore from the
boundary and starting conditions.
{The \fR -relations derived  by \citet{lopez2014} and our MESA
model without atmospheric escape have the similar shape throughout
the evolution, but the former predicts an atmospheric mass that is
smaller by less than 30\% (within our set of simulated planets)
than the latter. Therefore,  adding atmospheric escape to
\citet{lopez2014}, see the Lopez+HBA model in Figure
\ref{fig::rpl_fats_intro} (dotted light blue line), leads to a
similar systematic shift in atmospheric mass to our \fR -relation
(black solid line).}

\citet{lopez2014} considered the planetary radius \Rpl\ as the sum
of the core radius (${R_{\rm core}}/{R_{\oplus}} \approx ({M_{\rm
pl}}/{M_{\oplus}})^{0.25}$) and the radius of the convective
hydrogen-helium dominated envelope ($R_{\rm env}$), and the radius
of the radiative upper part ($R_{\rm atm} \approx 9{k_{\rm
b}T_{\rm eq}}/({g\mu_{\rm H/He}})$), where the latter is estimated
assuming the pressure at the photosphere of 20~mbar and at the
radiative-convective boundary of 100~mbar. $R_{\rm env}$ is given
by the analytical approximation based on 1300 thermal evolution
models and depend on the planetary mass, atmospheric mass
fraction, incident stellar flux and age of the planet.

As the photosphere in our MESA runs is defined in a different way
and normally lies {at a lower altitude than that given at
20~mbar}, we consider in addition to the total \Rpl\ given by this
approximation (light blue dotted line in
Figure~\ref{fig::rpl_fats_intro}), the reduced value of $(R_{\rm
core} + R_{\rm env})$ only, which gives on average  about 10\%
smaller radii than the ``full radius'' model (dark blue dotted
line). {Comparing the general properties of ``full radius'' and
``reduced radius'' model (i.e., without $R_{\rm atm}$), we see
that the inclusion of $R_{\rm atm}$ on the top of convective zone
barely changes the atmospheric mass for this specific planet. This
is because the removed upper layers have low atmospheric density.
However, the radius changes considerably and is about 10-15\%
lower for the ``reduced radius'' models shown  in
Figure~\ref{fig::rpl_fats_intro}, when comparing models at the
same age.} {The pressures at the upper boundary for the ``full
radius'' model and for the ``reduced radius'' model are  20~mbar
and 100~mbar, respectively. The pressure at the upper boundary
given by MESA is in general between these values for the planets
considered in this paper (in particular, for the planet considered
in this section it remains $\sim$76~mbar throughout the
evolution). Therefore, if the coupling between thermal evolution
and atmospheric escape is of minor importance,} we expect the
planetary radius given by MESA for the specific planetary
parameters to lie in between these two estimations for most of the
cases (except for the lightest planets in the present set, where
the radii at the young ages are strongly affected by the hotter
start in \citealt{lopez2014}). However, as one can see in
Figure~\ref{fig::rpl_fats_intro}, for the 12.5\Me -core planet,
this remains true only for the first $\sim$100~Myr, presumably due
to the inclusion of mass loss. This behaviour is similar for the
other planets in our framework, and we investigate this in more
detail in Section~\ref{ssec::fat_all}.

At last, we consider the estimation of the atmospheric mass
fraction given by the model of \citet{johnstone2015} (violet line
in Figure~\ref{fig::rpl_fats_intro}). This model is based on the
TAPIR (The AdaPtive, Implicit RHD) code developed by
\citet{steokl2015}, which provides radiative hydrodynamical models
for the atmospheres of planets with a given mass, radius and
equilibrium temperature. The nebula-accreted atmosphere in this
case is considered to be in hydrostatic and thermal equilibria,
and the luminosity of the core (established at the formation) is
the main heating source. The model employs the EOS by
\citet{saumon1995}, gas opacities by \citet{freedman2008} and dust
opacities by \citet{semenov2003}. As this model is initially
intended to describe the atmospheres newly accreted from the
protoplanetary disk, its natural upper boundary is the Hill
radius. For the estimates of the atmospheric mass fractions we
define the photospheric pressure as in \citet{kubyshkina2018grid}.
We exploit the grid of atmosphere models given by this approach
covering the same range of planetary parameters as the grid of
upper atmosphere models \citep{kubyshkina2018grid} used to
construct HBA, and interpolate within this grid to estimate the
atmospheric mass fractions.

Though this model employs the same EOS and the same gas opacities as MESA, there are several differences in the modelling approaches. 
For the planet considered in Figure~\ref{fig::rpl_fats_intro},
this model gives an estimate similar to the one given by the
present model (compare violet and solid black lines). The
difference in the atmospheric masses for the given radius
throughout the evolution remains on average about $0.04M_{\rm pl}$
(i.e., about 20\% of the initial envelope). The S-shaped variation
in the middle of the track corresponds to the variation in the
stellar temperature (and therefore, heating of the planet)
predicted by MIST. The temperature gradient in the model of
\citet{steokl2015,johnstone2015} is determined from radiative
transport whereas in MESA, the temperature gradient is found by
combining radiative transport with mixing-length theory. As a
result, the different treatments makes the atmosphere more
sensitive to heat variations in the model of
\citet{steokl2015,johnstone2015}. {This model also does not
account for the thermal evolution of the planet, so in general the
difference of this estimate to the results of MESA+HBA is expected
to be the largest.}

Overall, for all semi-analytical prescriptions, the resulting
radii at the end of the evolution are similar to those given by
the self-consistent modelling of the thermal evolution and
atmospheric mass loss (within $\sim$20\% uncertainty), but the
corresponding atmospheric mass fractions can differ significantly.
All the models shown in Figure~\ref{fig::rpl_fats_intro} predict a
final radius of $\sim$3-3.4~\Rer, but the atmospheric mass
fraction ranges from 0.025 to 0.06 \Mpl. {Next, we perform the
same comparison for all simulated planets in our sample.}

\subsubsection{Comparison for all planets}\label{ssec::fat_all}
 For all of the simulated planets considered in this work we see a similar behavior to what we saw in the case of the 12.5\Me -core planet. Differences in \fat\ increase with age for the approaches of \citet{johnstone2015} (violet) and \citet{lopez2014} with ``full radius'' (light blue) up to about 10 times (with an average increase of 2.5 times), while for the approach of  \citet{lopez2014} with ``reduced radius'' (dark blue),  \fat\  decreases from {$\Delta$\fat\ $\simeq$} 10-100\% {\fat\ }
to {$\Delta$\fat\ $\simeq$} 4-20\%{\fat\ .} The differences in
\fat\ given for a specific \Rpl\ by different approaches increase
with age by one order of magnitude for all three prescriptions for
{\fat\ $\lesssim$} 1\%; also, the difference given by the
approximation of \citet{johnstone2015} (not accounting for the
thermal evolution) increases relative to the others, while for the
first $\sim$100~Myr they remain at the same average level.

\begin{figure}
  \includegraphics[width=\hsize]{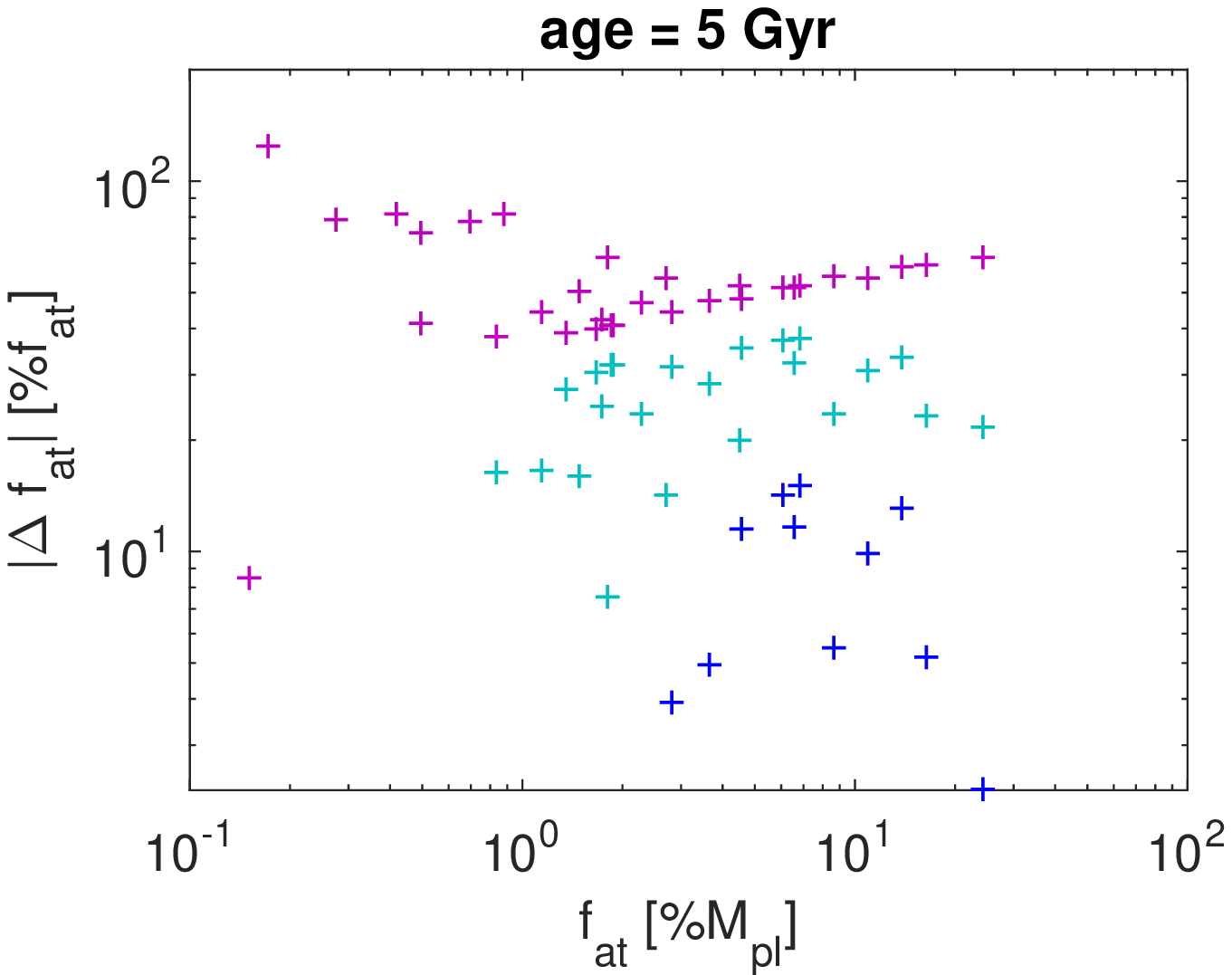}\\
  \caption{Relative differences between our preferred model (MESA+HD) and the three other models presented in Figure~\ref{fig::rpl_fats_intro} at an age of 5~Gyr. As in Figure~\ref{fig::rpl_fats_intro}, light blue color corresponds to the approximation from \citet{lopez2014}, dark blue is for the same approximation, but without the upper radiative part of the atmosphere, and magenta is for estimates based on \citet{johnstone2015}.}\label{fig::diff}
\end{figure}

We summarize  the difference between models for all our simulated
planets  in Figure~\ref{fig::diff}, where we plot the relative
difference in the atmospheric mass ($100\%\times\Delta f_{\rm
at}/f_{\rm at}$) given by the semi-analytical approaches to the
one given by MESA+HBA. The color coding remains the same as in
Figure~\ref{fig::rpl_fats_intro}. We consider this distribution at
5~Gyr for two reasons. First, the difference in the thermal state
of the planet for the models considering and not considering mass
loss reaches its maximum at the end of simulation, therefore,
these discrepancies can be considered as upper limits. Second,
this allows one to minimize the implicit difference in \fR\
through time caused by the different starting conditions in
\citet{lopez2014} compared to our approach (they assume the hotter
start, i.e., the initial entropy of the atmosphere is slightly
higher).

Considering the four different modelling approaches at 5~Gyr, for
\fat $\gtrsim2\%$, for most of the simulated planets the absolute
value of $\Delta f_{\rm at}/f_{\rm at}$ varies between 4 and 70\%,
where the upper limit is outlined by the approximation of
\citet{johnstone2015} (magenta points). {Overall, the
discrepancies of \citet{johnstone2015} with our reference model
are larger than those of other models, but they remain at a
similar level at different ages, which can make it a better
approximation for the most compact atmospheres, since systematic
differences are easier to correct.}

Because the differences between our coupled model (MESA+HBA) and
the three other approximations are usually below 70\% for $f_{\rm
atm} \gtrsim 2\%$, these four approximations give acceptable
estimates of \fR\ at 5~Gyr. However, if one requires a detailed
characterisation of the planet, we conclude that treating the
thermal evolution of the planet self-consistently with the
hydrodynamical modelling of atmospheric escape should be
considered, in particular for the low mass planets.

\section{Discussion and conclusions}\label{sec::discussion}

{Here, we modelled the long-term evolution of sub-Neptune planets
orbiting solar-like stars. Our models considered planets with a
range of core and atmospheric masses. We used MESA to model the
thermal evolution of planetary atmospheres and self-consistently
included atmospheric escape in our simulations. MESA has been used
before to study planetary evolution, but the novelty of our work
is that we used the more realistic prescription for atmospheric
escape from hydrodynamical studies
\citep{kubyshkina2018grid,kubyshkina2018approx}. In particular, we
incorporated the `hydro-based approximation' (HBA, Equations
\ref{eqn::HBA}), which is an analytical approximation from grids
of escape models, into MESA. Our inlists and supplement functions
can be accessed from the MESA market.  }

We found that the relation between the atmospheric mass fraction
and the radius of the planet is controlled to a large extent by
the thermal state of the atmosphere, which changes with time as
the planet cools down after formation, and can be significantly
affected by atmospheric escape. For the set of planets considered
in here, we found that the difference in temperature at the bottom
of the atmosphere for the escaping and non-escaping atmospheres
can reach a few thousand kelvin after 5~Gyr. This implies that
atmospheric escape changes the thermal gradient in the atmosphere,
with the effect being larger for the smaller planetary masses with
expanded atmospheres. In turn, the variations in the thermal state
of the planet (and therefore, the radius) affects the escape rate.
This implies that an accurate and self-consistent prescription of
atmospheric mass loss is of crucial importance in the study of
atmospheric evolution, in particular for low mass (below
$\sim$15\,\Me) planets.

Despite the differences caused by atmospheric escape, there are
some general trends that are similar for the cases of escaping and
non-escaping atmospheres. In both cases, the atmospheric mass
fraction corresponding to a specific planetary radius changes with
time as ${\rm age}^{0.11}$. The dependence on planetary mass
retains qualitatively the same behaviour in both cases as well;
however, quantitatively it differs {more} for low planetary masses
and atmospheric mass fractions above $\sim$10\%.

{There has been other studies that self-consistently incorporated
escape in MESA \citep[e.g.,][]{Chen_rog2016}, albeit using a
different prescription for planetary mass loss than the HBA. To
compare with previous works, we also  simulated thermal evolution
of planets using the commonly adopted energy limited (EL)
approximation.} Concerning the planetary evolution, the HBA
prescription leads to higher chances of survival of the primary
atmospheres for more compact planets, compared to what was
obtained applying the EL approximation. Furthermore, the HBA
prescription moves the boundary between old planets with and
without a primary atmosphere towards higher mass planets.

We also compared our results with other models, such as that of
\citet{lopez2014}, which considers a different thermal evolution
model from that given by MESA without atmospheric escape, and that
of \citet{johnstone2015}, which {calculated atmospheric mass
without} accounting for thermal evolution. {To make the comparison
more adequate, we implemented the HBA mass loss on these models
and compare their results with our MESA+HBA models.} In {all
these} cases, the planetary radii at the end of the simulation
(5Gyr) are relatively similar, with a difference below 20\%, to
those predicted by the combination of MESA with HBA. The
atmospheric mass fractions, however, {show a difference of up to
70\% compared to our models for planets with atmospheric mass
fractions above 2\%, and up to a few times for the lighter
atmospheres.} This can not be explained only by differences in the
approaches used in the models, as the differences in \fR\ relation
grow systematically with age; therefore, the larger differences
{compared} to the case of non-escaping atmospheres should come
from the effect on the thermal evolution caused by atmospheric
mass loss. From this, we concluded that, separately considering
atmospheric escape and thermal evolution can yet be acceptable for
rough estimates of planetary radii and atmospheric mass fraction
for larger atmospheres, or for massive planets with relatively
compact envelopes.

However, for an accurate modelling of the atmospheric escape or
characterisation of  observed exoplanets, the thermal evolution
process and atmospheric escape should be considered
self-consistently. Our findings show that the choice of a more
realistic atmospheric mass loss prescription is crucial for
evolutionary studies. New instrumentation, such as the CHEOPS
satellite \citep[][]{broeg2013}, which are capable of measuring
planetary radii to the 2-3\% level, will provide the necessary
accuracy enabling one to identify the effects of self-consistently
accounting for both planetary thermal evolution and mass loss.

\section*{Acknowledgements}
This project has received funding from the European Research
Council (ERC) under the European Union's Horizon 2020 research and
innovation programme (grant agreement No 817540, ASTROFLOW).

\section*{Data Availability}
The data underlying this article are available in Zenodo
Repository, at  https://doi.org/10.5281/zenodo.4022393.







\appendix
\section{Effective radii of the stellar XUV absorption.}\label{apx::Reuv}

\begin{figure}
  \includegraphics[width=\hsize]{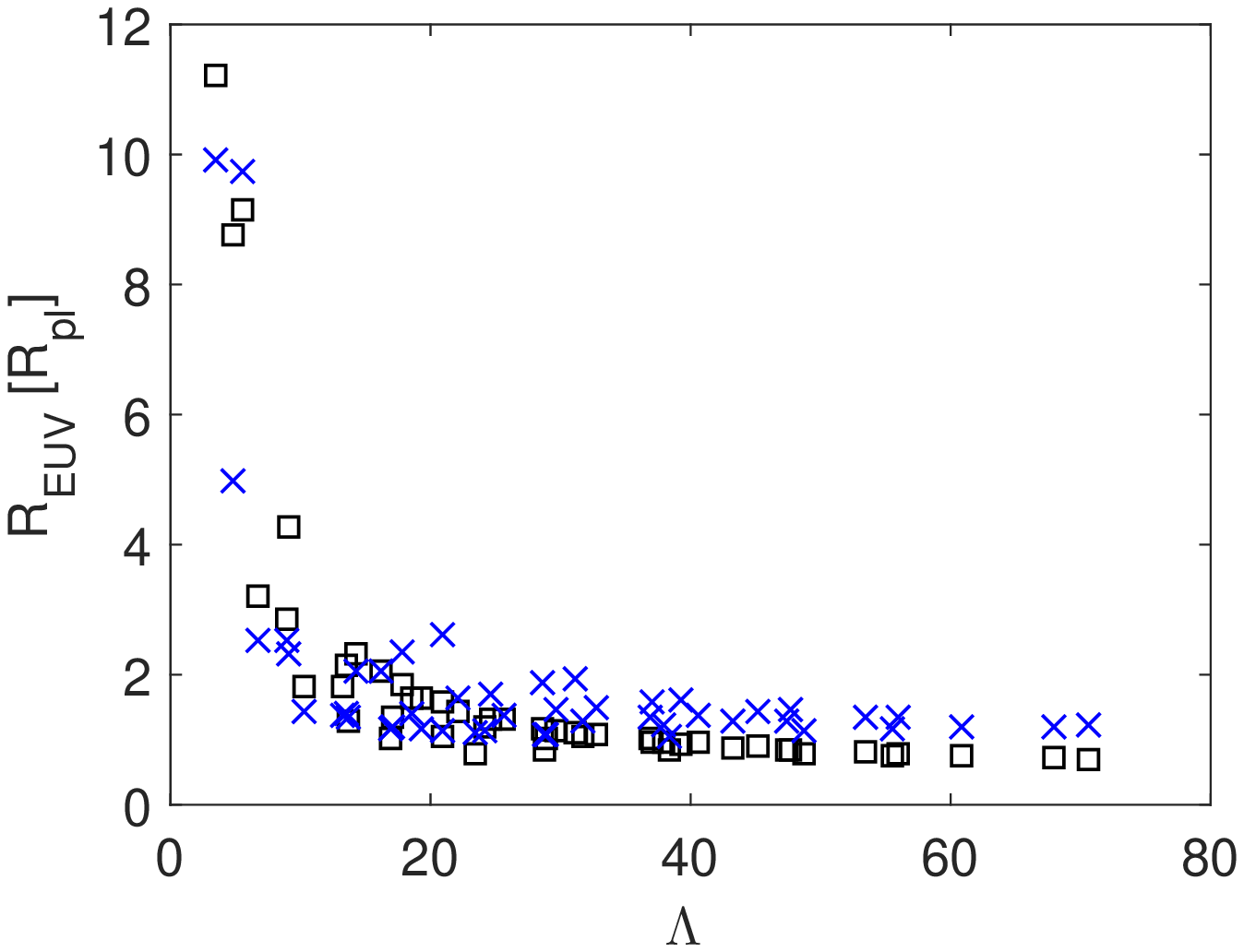}\\
  \caption{Effective radius of the absorption of stellar XUV radiation for the sample of planets considered in this work. Black squares represent the radii used for the energy limited approach (as given by Equation~\ref{eq::Reuv}), and blue crosses show the $R_{\rm XUV}$ obtained for the same planetary parameters by interpolation within the grid of hydrodynamical models of upper atmospheres \citep{kubyshkina2018grid}.}\label{fig::Reuv}
\end{figure}

Here we present the distribution of the effective radii of the
stellar XUV absorption predicted by the Equation~\ref{eq::Reuv}
for the planets in our data set in comparison to the predictions
of HD models (Figure~\ref{fig::Reuv}). The latter were obtained
through the interpolation within the grid of upper atmosphere
models presented in \citet{kubyshkina2018grid}. The two estimates
agree well for the gravitational parameters $\Lambda\gtrsim15$.

\section{Cubic polynomial fits coefficients}\label{apx::cfu}

In Table~\ref{tab::cfu} we present the coefficients of the cubic
polynomial approximations of \fR\ dependence for all the planetary
core masses considered in this work at different moments in time.
These coefficients are of the restricted use, as they are bound to
the specific system configuration. They can, however, be used to
test the general trends in \fR .

\begin{table*}\label{tab::cfu}
\caption{Here we present the coefficients of the cubic polynomial
approximation $f_{at} = c_0 + c_1R_{pl} + c_2R_{pl}^{2} +
c_3R_{pl}^{3}$ for the set of planets simulated in the present
work at different ages. Note that these coefficients are of the
restricted use, as they are bound to the specific system
configuration. }
\begin{tabular}{|c|c|c|c|c|c|}
  \hline
  $M_{\rm core}$ [$M_{\oplus}$] & age [Myr] & $c_0$ & $c_1$ & $c_2$ & $c_3$ \\
  \hline
  7.5 & 6    & -9.5174186e-02 &  3.7805986e-02 &  2.4897235e-03 & -5.4846104e-05 \\
  7.5 & 10   & -9.6726476e-03 & -1.5654047e-02 &  1.1662104e-02 & -5.5184314e-04 \\
  7.5 & 30   & 1.1613491e-02  & -4.0110695e-02 &  1.9335906e-02 & -1.2887226e-03 \\
  7.5 & 50   & 3.3928231e-02  & -6.2786585e-02 &  2.6864042e-02 & -2.0914725e-03 \\
  7.5 & 100  & 8.2414444e-02  & -1.1580330e-01 &  4.5702243e-02 & -4.2533551e-03 \\
  7.5 & 500  & 2.4745440e-01  & -3.0179365e-01 &  1.1367423e-01 & -1.2295493e-02 \\
  7.5 & 1000 & 2.3423878e-01  & -2.8697032e-01 &  1.0822829e-01 & -1.1659277e-02 \\
  7.5 & 5000 & 3.2817317e-01  & -4.1002648e-01 &  1.6175864e-01 & -1.9374941e-02 \\
    &   &   &   &    &   \\
  10  & 6    & -8.4337932e-02 &  2.1263943e-02 &  7.2651128e-03 & -2.5831732e-04 \\
  10  & 10   & -1.1829657e-01 &  4.7280208e-02 &  1.2402145e-03 &  1.2094692e-04 \\
  10  & 30   & -2.0784025e-02 & -1.4012184e-02 &  1.2077970e-02 & -5.2396709e-04 \\
  10  & 50   & -8.2177727e-03 & -2.6373716e-02 &  1.5658310e-02 & -8.0826119e-04 \\
  10  & 100  & -1.5884565e-02 & -2.3379397e-02 &  1.5859988e-02 & -8.8580159e-04 \\
  10  & 500  &  6.4912519e-02 & -9.7909313e-02 &  3.7497361e-02 & -2.7851885e-03 \\
  10  & 1000 &  1.0935944e-01 & -1.3935513e-01 &  4.9605804e-02 & -3.8843362e-03 \\
  10  & 5000 &  9.2015409e-02 & -1.2035197e-01 &  4.2755580e-02 & -3.1194446e-03 \\
    &   &   &   &    &   \\
  12.5 & 6    & -1.0294407e-01 &  2.9212775e-02 &  6.0305936e-03 & -1.3372028e-04 \\
  12.5 & 10   & -9.8722962e-02 &  2.8063438e-02 &  5.8132746e-03 & -1.2044831e-04 \\
  12.5 & 30   & -8.9115448e-02 &  2.8857909e-02 &  3.5865897e-03 &  6.9361893e-05 \\
  12.5 & 50   & -5.8511717e-02 &  8.4950426e-03 &  7.6332101e-03 & -1.6251869e-04 \\
  12.5 & 100  & -3.7070666e-02 & -8.1771528e-03 &  1.1563853e-02 & -4.2972074e-04 \\
  12.5 & 500  &  7.5268361e-03 & -4.8910663e-02 &  2.2855972e-02 & -1.3199504e-03 \\
  12.5 & 1000 &  4.7302452e-02 & -8.2469558e-02 &  3.1634502e-02 & -2.0195481e-03 \\
  12.5 & 5000 &  9.5876225e-02 & -1.2094996e-01 &  4.1060868e-02 & -2.7550134e-03 \\
    &   &   &   &    &   \\
  15  & 6    & -1.1683502e-01 &  3.9591844e-02 &  2.7564329e-03 &  1.6832543e-04 \\
  15  & 10   & -1.3013691e-01 &  5.1074214e-02 & -2.0892143e-04 &  3.6535495e-04 \\
  15  & 30   & -7.7774615e-02 &  1.7816419e-02 &  5.9227288e-03 & -2.9717993e-05 \\
  15  & 50   & -8.9808994e-02 &  2.8189577e-02 &  3.2041033e-03 &  1.9096933e-04 \\
  15  & 100  & -8.2775974e-02 &  2.3982909e-02 &  3.8103128e-03 &  1.9005829e-04 \\
  15  & 500  & -5.5353478e-02 &  2.1545099e-03 &  8.8578143e-03 & -1.0428678e-04 \\
  15  & 1000 & -2.3641070e-02 & -2.2115588e-02 &  1.4460258e-02 & -4.8270803e-04 \\
  15  & 5000 &  3.8668347e-02 & -6.8920312e-02 &  2.5215896e-02 & -1.2478985e-03 \\
    &   &   &   &    &   \\
  20  & 6    & -2.6870843e-02 & -2.7392640e-02 &  1.6457224e-02 & -6.1203891e-04  \\
  20  & 10   & -2.3393905e-02 & -2.8632278e-02 &  1.6496182e-02 & -6.2818946e-04  \\
  20  & 30   & -1.2478324e-02 & -3.3819665e-02 &  1.7023745e-02 & -6.8905757e-04  \\
  20  & 50   & -1.4195873e-02 & -3.2968626e-02 &  1.6970950e-02 & -6.9119069e-04  \\
  20  & 100  & -1.4581100e-02 & -3.3055789e-02 &  1.7057524e-02 & -6.9844106e-04  \\
  20  & 500  &  3.2123656e-03 & -4.9012298e-02 &  2.1278458e-02 & -1.0175440e-03  \\
  20  & 1000 &  3.3072808e-02 & -7.1569615e-02 &  2.6548903e-02 & -1.3996529e-03  \\
  20  & 5000 &  8.4434684e-02 & -1.0810294e-01 &  3.4681657e-02 & -1.9969340e-03  \\
  \hline
\end{tabular}
\end{table*}

%


\bsp    
\label{lastpage}
\end{document}